\documentclass[acus]{JAC2003}

\pdfoutput=1
\usepackage{graphicx}
\usepackage{booktabs}
\begin{document}
\title{MICE Particle Identification Systems}

%\author{D. A. Sanders, L. M. Cremaldi and D. J. Summers, \\
%University of Mississippi, University, MS 38677, USA (for the MICE Collaboration) \\
\author{D. A. Sanders\thanks{for the Mice Collaboration}  \\ 
University of Mississippi, University, MS 38677, USA \\
}

\maketitle

\begin{abstract}
The international Muon Ionization Cooling Experiment (MICE) is being built, at the Rutherford Appleton Laboratory (RAL), to demonstrate the feasibility of ionization cooling of muon beams. This is one of the major technological steps needed in the development of a muon collider and a neutrino factory based on muon decays in a storage ring.  MICE will use particle detectors to measure the cooling effect with high precision, achieving a precision on the measurement of emittance of 0.1\% or better.  The particle i.d.  detectors and  trackers must work in harsh environmental conditions due to high magnetic fringe fields and RF noise.  We will briefly describe the MICE particle i.d. detector systems, and show some current performance measurements of these detectors.
\end{abstract}

\section{Introduction}

Cooled muons are required for a Neutrino Factory based on a muon storage ring \cite{Ayres}, \cite{Blondel}, \cite{Albright}, \cite{Berg} and for a muon collider \cite{Neuffer}, \cite{Ankenbrandt}, \cite{Summers}, \cite{Palmer}.  The Muon Ionization Cooling Experiment (MICE) \cite{Drumm}, \cite{Yoshida} at Rutherford-Appleton Lab is the first test of the ionization cooling concept for muon beams in the approximate momentum range 140 to 240 MeV/c. 
A minimum ionizing muon beam will be transversely cooled by stages of $dE/dx$ in LH$_2$ absorbers and longitudinal energy restoration in a series of 201 MHz RF cavities; see Figure 1.
The 6D emittance reduction is measured before and after the cooling stage by tracking individual muons through the system. To establish muon cooling the in-flight muon beam is positively identified  by three time-of-flight (TOF) stations \cite{Bonesini}, by two threshold Cherenkov detectors (CKOVs), and by a low energy ranging electron-muon calorimeter (KL/EMR) near the  beam exit.

\begin{figure}[htb]
   \centering
   \includegraphics*[width=80mm]{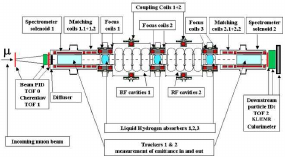}
   \caption{MICE Spectrometer Layout.}
   \label{fig1-spectrometer}
\end{figure}
\vfill

\section{Time of Flight System}
 
Three time-of-flight (TOF) stations are positioned in the MICE channel at the beginning (TOF0), midway (TOF1), and near the rear (TOF2). TOF0, TOF1 and TOF2 are respectively 40 x 40 cm$^2$,
42 x 42 cm$^2$ and 60 x 60 cm$^2$ in active cross section and spaced apart by a $\approx$10 m flight path.  The TOF 
stations are used in establishing a precision particle trigger which can be synchronized to within $\leq 70$ ps of the RF cavity phase. 

\begin{figure}[htb]
   \centering
   \includegraphics*[width=70mm]{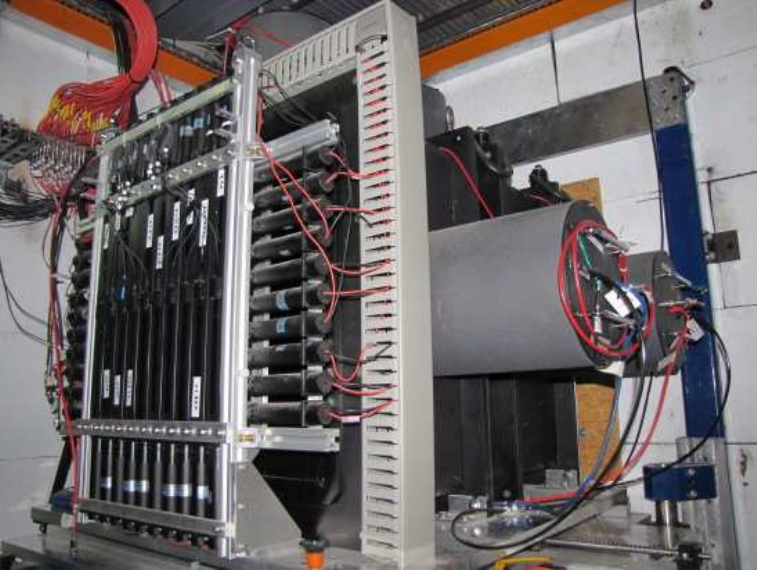}
   \caption{X/Y planes for TOF0 station. Each bar assembly is 4 cm wide. (The CKOV detectors are behind the TOF0)}
   \label{figure3-tof0_xy}
\end{figure}

The TOF 0/1/2 stations consist of 10/7/10 X-counter  and 10/7/10 Y-counter arrays constructed of BC404/420/420 scintillator bars with dual R4998 PMT readout, see Figure \ref{figure3-tof0_xy}. The HV dividers have been modified for high rate performance ($\approx$2 MHz).  The dual photomultiplier (PMT) readout gives typically $\sigma_t$=50-60 ps intrinsic timing resolution for each bar assembly. The bars are one inch thick, optimizing between light collection and energy loss.  

The transit time and associated dispersion, $\sigma_{tt}$, of the signal through the PMT, cable delay, and discriminating electronics is not known and are measured for each channel by a calibration procedure which can use particle beam and/or cosmic rays. A system based on a fast laser will be used to aid in TOF calibration and to track small timing changes. 

Leading edge discriminators have been adopted for the timing measurements.  
The adoption of these discriminators introduces a dependence of the discrimination crossing time, ``time-walk", with its associated dispersion $\sigma_{tw}$.  To calculate the time-walk correction, the difference of the time measured by the PMTs and a reference time is measured for a series of data. The correction function is then applied offline. 
The expected resolution of the time-of-flight is around 75 ps, including calibration errors.
\vfill\break

The TOF data acquisition utilizes three TDC boards (CAEN V1290). The digital values recorded by the TDCs correspond to the absolute time since the last reset of each TDC board.  A ``particle trigger" signal is generated by hardware logic units in the data acquisition (DAQ) racks. It is given by the first dual coincidence of the PMTs connected to the same TOF0 bar unit. The first channel of each TDC board receives a copy of the particle trigger signal and this signal is used as a reference for all the PMT signals of the TOF stations.

\begin{figure}[ht]
  \begin{minipage}{40mm}
   \centering
   \includegraphics[width=37mm]{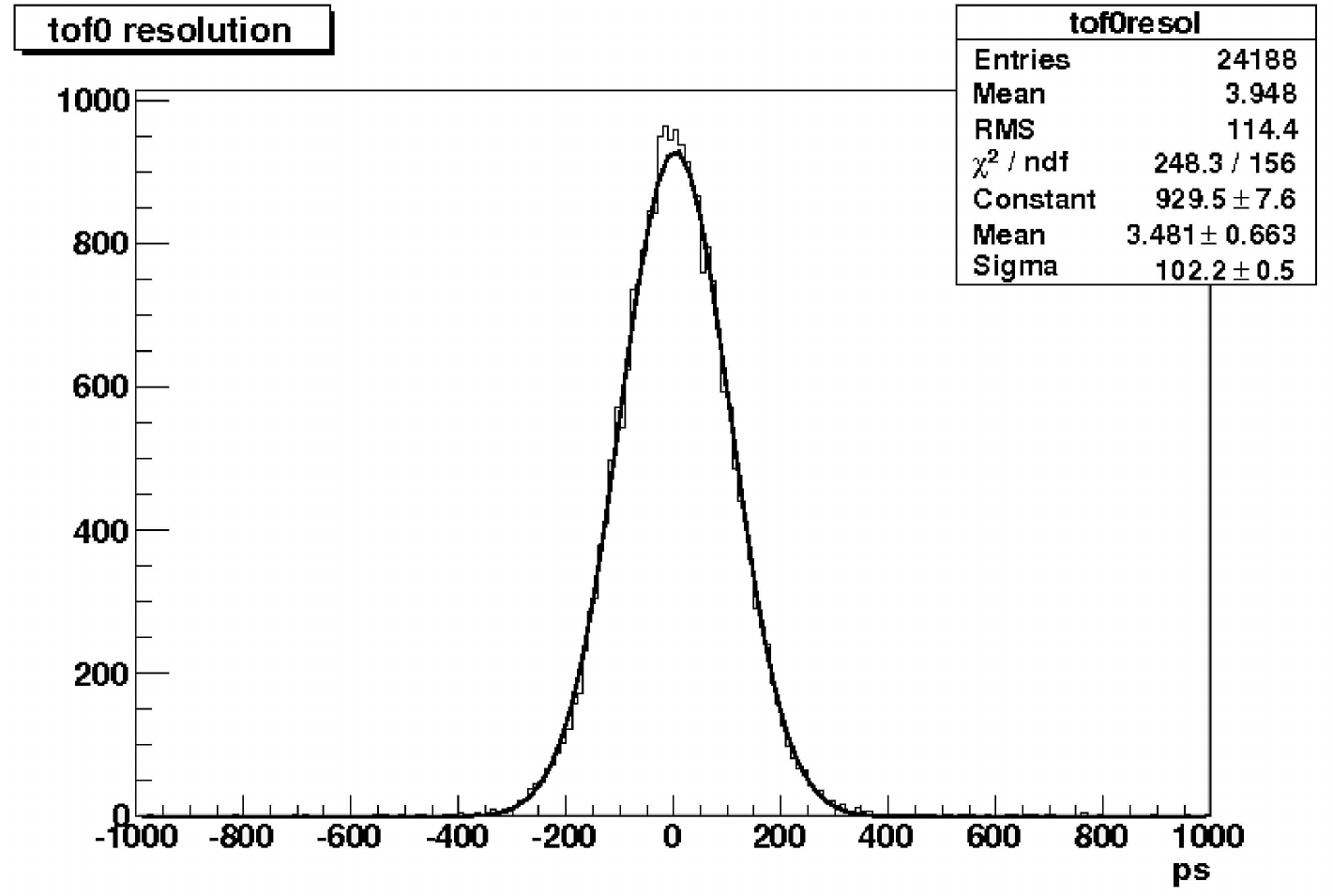}
   \caption{TOF0 detector's\,\,\,\,  intrinsic timing.}
    \label{figure3-tof0_res}
  \end{minipage}
% \hspace{2mm}
  \begin{minipage}{40mm}
   \centering
   \includegraphics[width=37mm]{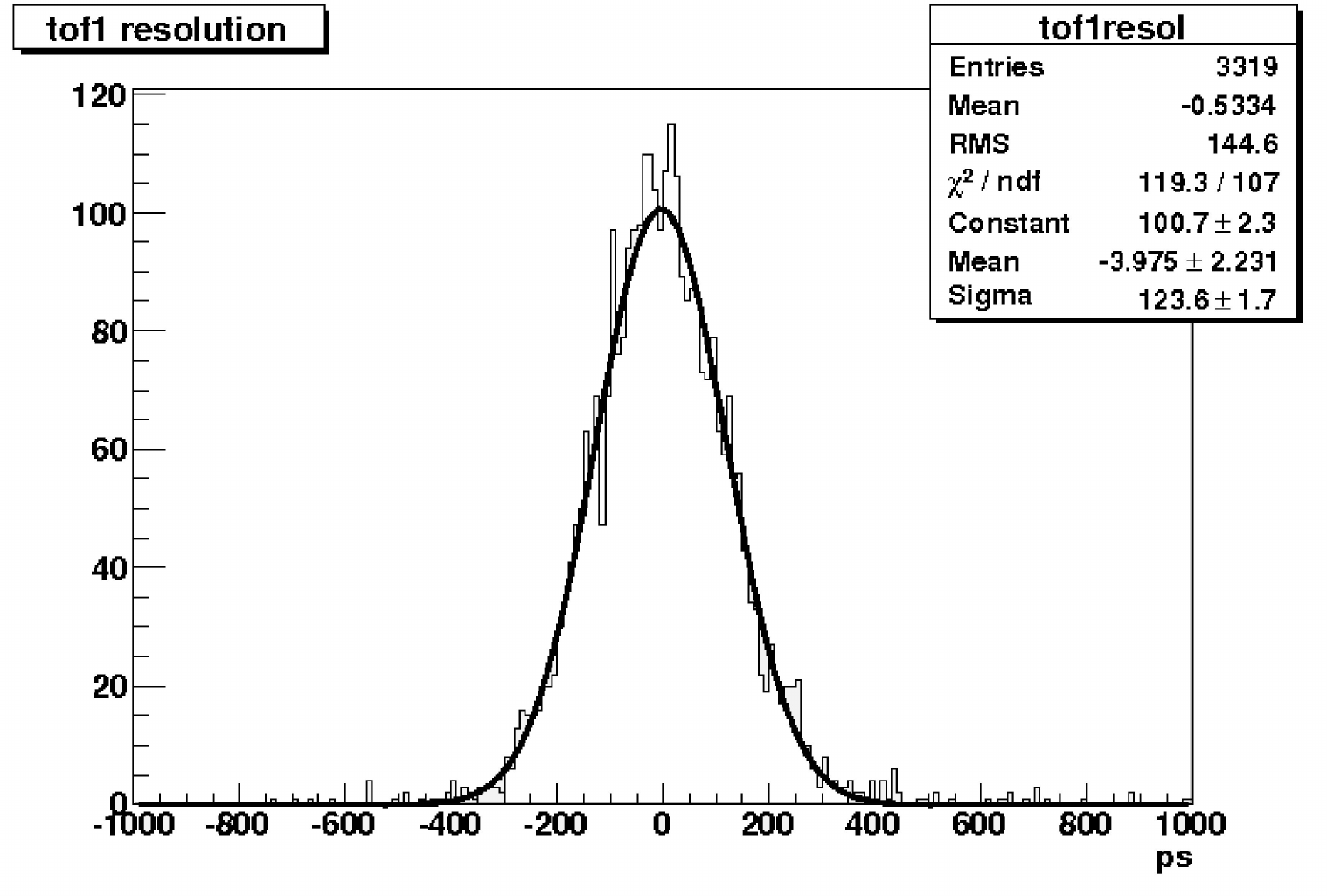}
   \caption{TOF1 detector's\,\,\,\,  intrinsic timing.}
   \label{figure4-tof1_res}
  \end{minipage}
\end{figure}

\begin{figure}[htb]
   \centering
   \includegraphics*[width=75mm]{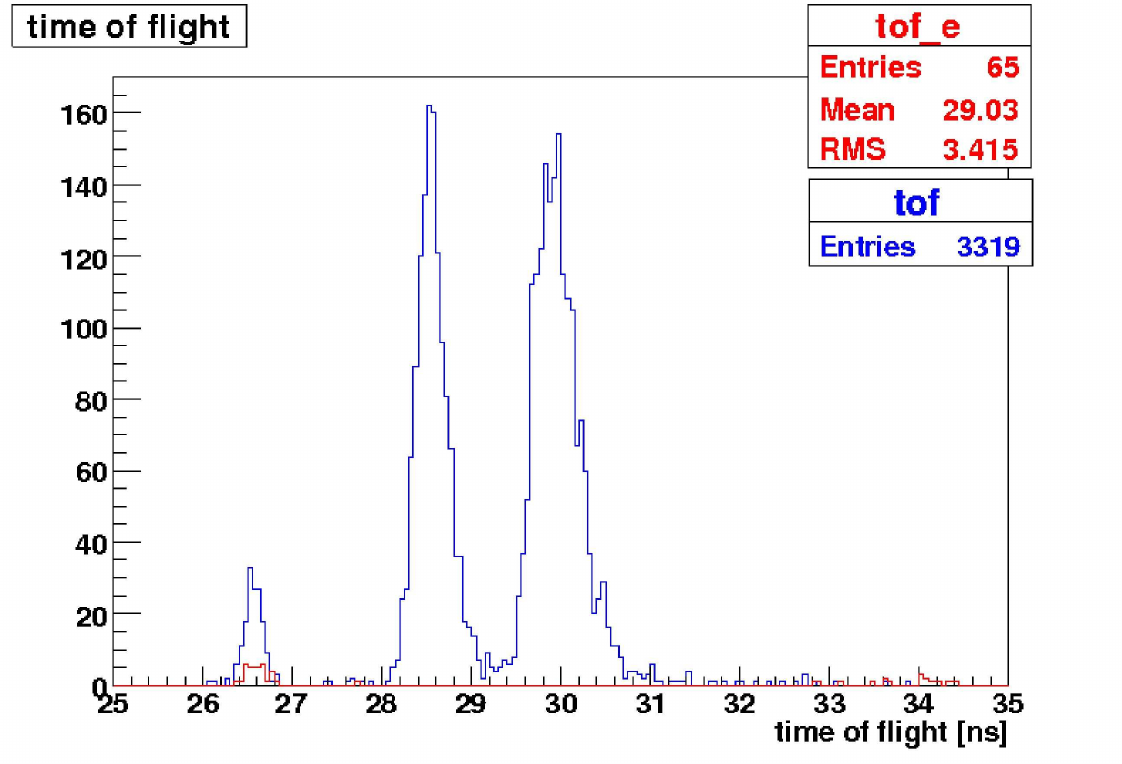}
   \caption{TOF1-TOF0 timing of $e^+,\mu^+,\pi^+$ in nominal  300 MeV/c $\pi^+$ beam.}
   \label{ figure5-tof01}
\end{figure}

In the fall of 2008, components of the TOF0 and TOF1 stations were commissioned in the MICE beamline. For a set of pion runs, with time-walk corrections applied, the intrinsic detector resolutions were measured to be in the 55 ps to 65 ps range, very close to design specification.  These intrinsic timing distributions are shown in Figure \ref{figure3-tof0_res} and Figure \ref{figure4-tof1_res}. The time-of-flight distribution for the nominal 300 MeV/c pion tune is shown in Figure \ref{ figure5-tof01}, displaying a  good separation between positrons, muons, and pions in the beam at this early stage of analysis. 

\section{Aerogel Cherenkov Counters}

Between 220 MeV/c and 360 MeV/c the time-of-flight difference  between $\mu$'s and $\pi$'s varies from about 2.4 ns to 1.0 ns over a 10 meter flight path. Two aerogel Cherenkov counters were designed to aid in the separation, with the emphasis placed on higher momenta. By using the  the two devices in  tandem 4-5$\sigma$ $\mu/\pi$ separation can be achieved over the full momentum range. The CKOV detectors are located upstream of the diffuser and thus are designed to operate in the 220 to 360 Mev/c rather than the 140 to 240 MeV/c  cooling channel momentum range. They were commissioned in the summer of 2008.

\begin{figure}[htb]
   \centering
   \includegraphics*[width=55mm]{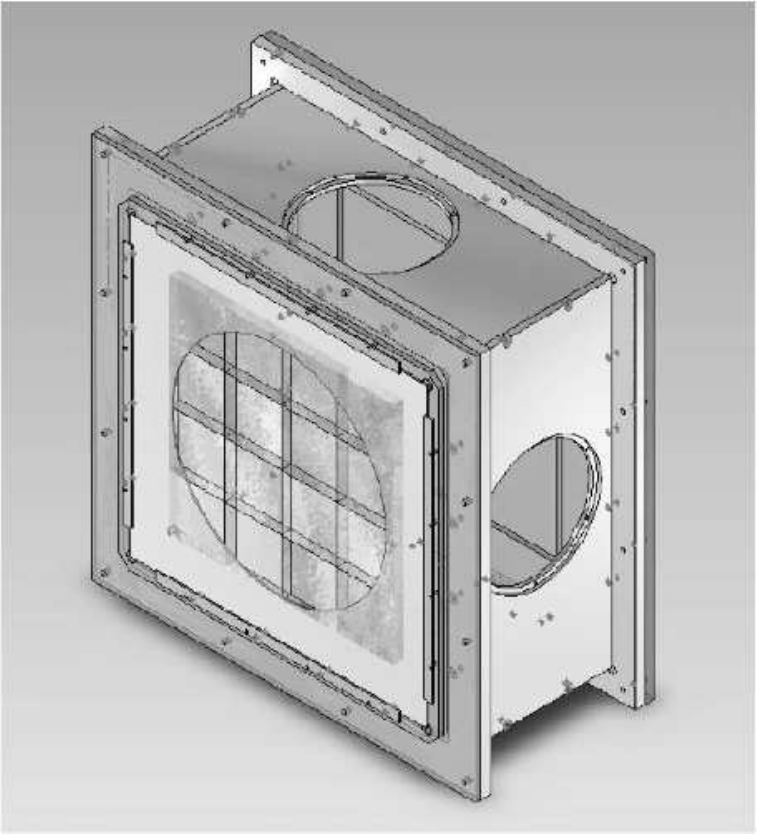}
   \caption{Model sketch of aerogel Cherenkov detector. }
   \label{ figure6-ckov_model}
\end{figure}

High density aerogel Cherenkov radiators were selected, \cite{sorma08},  with indices of refraction and muon threshold momenta of $n$=1.07 ($p_{th}^{\mu}$=278 MeV/c ) and $n$=1.12 ($p_{th}^{\mu}$=220 MeV/c).  Each counter consists of a 2.3 cm thick aerogel radiator sealed behind a thin UV-transmitting window. The Cherenkov light is collected by four 8-inch Electron Tubes 9354KB PMTs. The PMT dividers have been designed to operate at $\leq$ 5 MHz beam rate with transistorized last dynode stages. The charge is digitized by a 500 MHz flash ADC (CAEN V1721 500MS/s FADC) in the DAQ rack.  A model of the aerogel Cherenkov detector with open PMT ports is shown in Figure \ref{ figure6-ckov_model}. 

\begin{figure}[htb]
   \centering
   \includegraphics*[width=70mm]{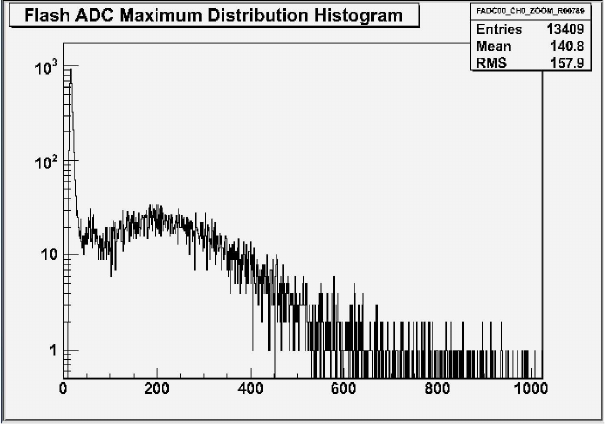}
   \caption{Typical light spectrum from single CKOV PMT, about 5-6 p.e. per PMT.}
   \label{ figure7-ckov_spec}
\end{figure}

The saturation ($\beta$=1) photoelectron calibration for particles is found in 200 MeV/c positron runs. A typical PMT spectrum is shown in Figure \ref{ figure7-ckov_spec}. We collect between 20 and 25 photoelectrons in the 4 PMTs of the aerogel counters. This yield gives sufficient photostatistics for a good muon tagging efficiency of 98\% at low pion misidentification rate ($\leq$10$^{-3}$~) over the 220-360 MeV/c momentum range.

\vfill

\section{KL and Electron-Muon Ranger }

Near the end of the final MICE tracking solenoid a state-of-the-art high-resolution electromagnetic Pb-scintillating fiber calorimeter (KL) and a rear electron-muon ranger (EMR) are positioned.  The KL are of the type built by KLOE \cite{Ambrosino:2009zza}. The KL/EMR makes the final clean muon tag vetoing $\mu^+\rightarrow e^+ \bar{\nu_e} \nu_{\mu}$ decays at the $10^{-2}$ level. The KL/EMR offers adequate energy resolution to perform muon and electron identification in the momentum range of interest.

The KL preshower calorimeter is constructed of 0.3mm grooved Pb with Bicron BCF-12 scintillating fiber inlay in a KLOE-type layout. The KL represents 2.5 $X_0$ in depth, a 4 cm thick active depth. It has an energy resolution for electrons  of ${\Delta E /E}  = 7\% / \sqrt{E}$(GeV) and timing resolution of $\Delta_t $ = 70 ps / $\sqrt{E}$. 

\begin{figure}[htb]
   \centering
   \includegraphics*[width=65mm]{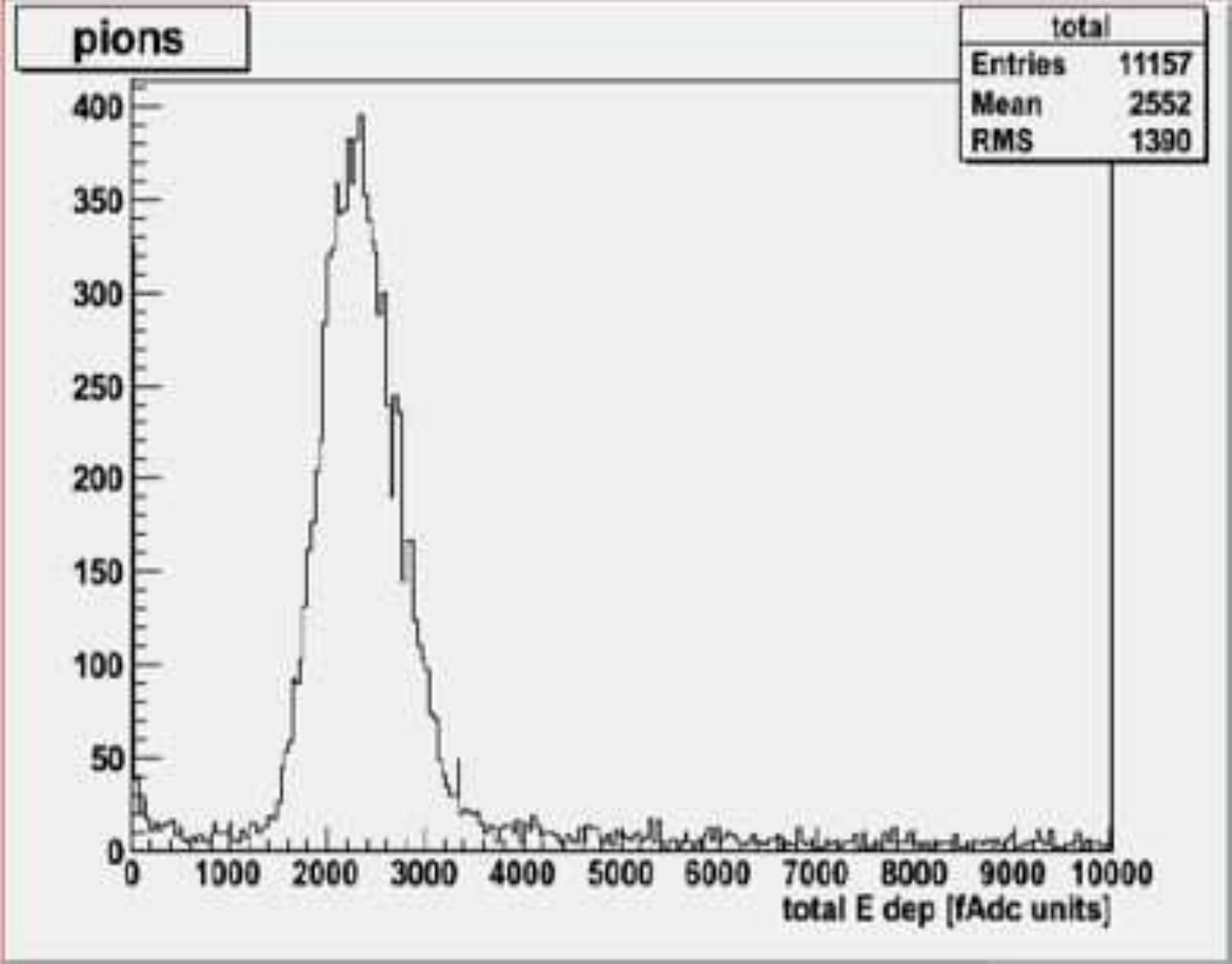}
   \caption{KL response to 300 MeV/c pion beam. The horizontal scale is in ADC channels.}
   \label{ figure8_kl}
\end{figure}

The KL was commissioned in the fall 2008 MICE run. Below we show a scan with 300 MeV/c pion beam in to the calorimeter. The ionizing response shown in Figure \ref{ figure8_kl} was as expected. Further calibrating exposures to electron beam are planned in 2010. 

\begin{figure}[htb]
   \centering
   \includegraphics*[width=45mm]{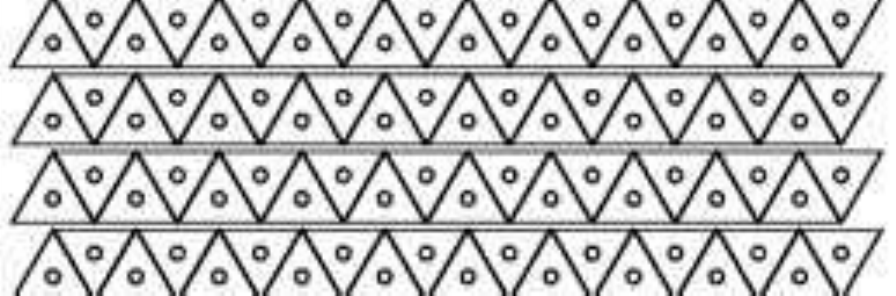}
   \caption{Sketch of MINER$\nu$A-like EMR calorimeter design to be used in MICE.}
   \label{figure9_minerva}
\end{figure}

\begin{figure}[htb]
   \centering
   \includegraphics*[width=65mm]{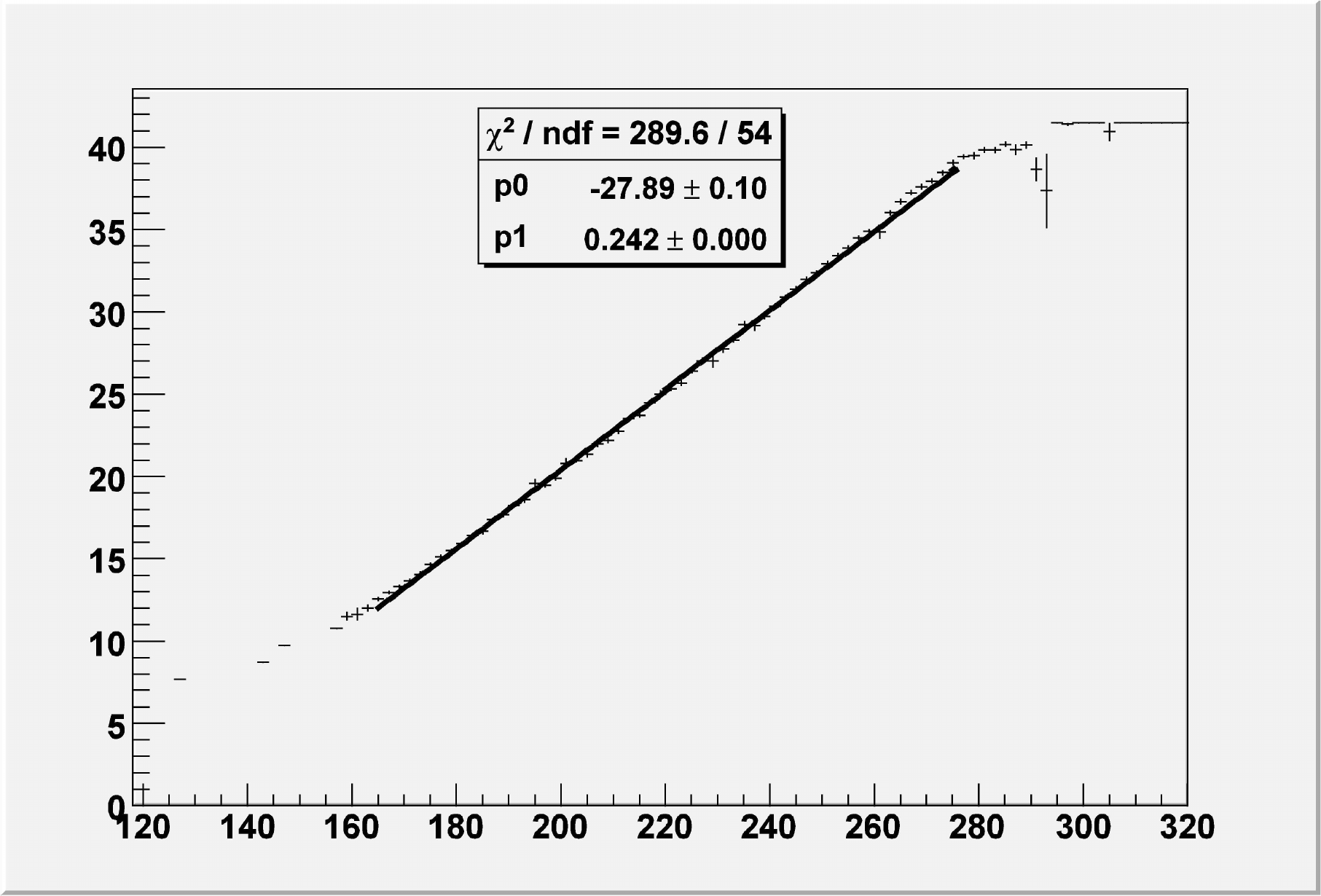}
   \caption{Muon range simulation in the EMR. $\sum{E}$ versus Layer \# is plotted. The horizontal axis is in MeV/c. Muons above 270 MeV/c will range through.}
   \label{figure10_range}
\end{figure}

Penetrating electrons and muons are further discriminated in the electron-muon ranger (EMR), an instrumented block of active scintillator, 68 cm deep. The EMR is currently under construction and utilizes a MINER$\nu$A-like design \cite{MINERvA}. The detector is arranged in 40 optically isolated layers of segmented readout. Each layer is composed of 59 triangle-shaped scintillator bars (Figure \ref{figure9_minerva})  fitted together in pairs. Each bar is read out through a central  WLS fiber (Bicron BCF-92 fiber, aluminized at far end). The scintillation light is piped to arrays of multi-anode PMTs. There are a total of 2360 readout channels. Simulations show that MICE can achieve a longitudinal momentum resolution of $\sigma_{P_z} \approx$ 4 MeV/c for muons under 270 MeV/c, which range out in the calorimeter (see Figure \ref{figure10_range}).  The error on the longitudinal momentum is comparable to that of the MICE tracker measurement. 

\end{document}